\documentclass[reprint,nofootinbib,amsmath,amssymb,fontenc aps]{revtex4-2}
\usepackage{graphicx,dcolumn,bm,hyperref,float}
\usepackage{anyfontsize}
\usepackage{color}

\newcommand{\beq}{\begin{equation}}
\newcommand{\eeq}{\end{equation}}
\newcommand{\bea}{\begin{eqnarray}}
\newcommand{\eea}{\end{eqnarray}}
\newcommand{\bef}{\begin{figure}}
\newcommand{\eef}{\end{figure}}

\newcommand{\f}{f_a}
\newcommand{\m}{m_a}
\newcommand{\Tqcd}{T_{\mbox{\tiny{QCD}}}}
\newcommand{\tauqcd}{\tau_{\mbox{\tiny{QCD}}}}
\newcommand{\gqcd}{g_{\mbox{\tiny{QCD}}}}
\newcommand{\mpl}{M_{\mbox{\tiny{Pl}}}}
\newcommand{\Ts}{T_*}
\newcommand{\mchi}{m_\chi}
\newcommand{\Sc}{S}
\newcommand{\Y}{Y}
\newcommand{\Fc}{F}
\newcommand{\LamDim}{\lambda}
\newcommand{\mudim}{\alpha}
\newcommand{\mratio}{\mu}

\newcommand{\Tchi}{T_\chi}

\begin{document}

\title{Altered Axion Abundance from a Dynamical Peccei-Quinn Scale}

\author{Itamar J.~Allali$^1$}
\email{itamar.allali@tufts.edu}
\author{Mark P.~Hertzberg$^1$}
\email{mark.hertzberg@tufts.edu}
\author{Yi Lyu$^2$}
\email{ylyu11@ucsc.edu}
\affiliation{$^1$Institute of Cosmology, Department of Physics and Astronomy, Tufts University, Medford, MA 02155, USA
\looseness=-1}
\affiliation{$^2$University of California Santa Cruz, Santa Cruz, CA, 95064, USA
\looseness=-1}

\begin{abstract}
We build a model in which the relic abundance of axions is altered from the standard misalignment mechanism, either increased or decreased, due to the presence of a new light scalar that couples to the radial part of the Peccei-Quinn (PQ) field. The light scalar makes the effective PQ symmetry-breaking scale dynamical, altering the early-time dynamics for the axion and affecting its late-time dark matter abundance. We analyze this new mechanism semianalytically and numerically, showing that we can accommodate both lighter or heavier axion dark matter, compared to the standard treatments. We discuss implications of the model for axion searches and fundamental physics.
\end{abstract}

\maketitle

\tableofcontents

\section{Introduction}

The strong CP problem in the Standard Model is the question of why the term $\Delta\mathcal{L}_{qcd}\sim\theta G\tilde{G}$ has $\theta\lesssim 10^{-10}$, an outstanding puzzle of fundamental importance. Its leading solution remains the introduction of a new (approximate) global Peccei-Quinn (PQ) symmetry \cite{Peccei1977}, 
which is spontaneously broken at some high scale, leaving an approximate Goldstone boson, the axion \cite{Weinberg1978,Wilczek1978}. The axion acquires a small, but nonzero, mass through QCD instantons. Its potential causes the field to head toward small values due to cosmic expansion. This leads to the effective CP violation in the strong sector being driven toward zero. 

As a by-product, the axion, which is electrically neutral and normally very stable, acts as a form of dark matter. Through the misalignment mechanism, wherein the axion's field is initially displaced from its minimum, it evolves to become nonrelativistic in the late universe, and contributes a relic dark matter abundance\cite{Preskill1983,Abbott1983,Dine1983}.

In minimal models of the QCD axion, with standard cosmological history, the relic abundance $\Omega_a\equiv\rho_a/\rho_{crit}$ is given approximately by
\beq
\Omega_a\sim \left(\f\over 10^{12}\,\mbox{GeV}\right)^{7/6}\langle\theta_i^2\rangle
\label{Omaxion}\eeq
where $\f$ is the PQ symmetry breaking scale and $\langle\theta_i^2\rangle$ is the spatially averaged value of the square of the initial displacement angle. So to get the observed dark matter abundance of $\Omega_a\approx 0.25$ (i.e., approximately 25\% of the universe today), one needs $\f$ to be a little smaller than $10^{12}$\,GeV (unless $\langle\theta_i^2\rangle$ is atypically small). The corresponding axion mass is given by
\beq
\m={\Lambda_0^2\over \f}\approx 5.7\,\mu\,\mbox{eV}\left(10^{12}\,\mbox{GeV}\over \f\right)
\eeq
(where $\Lambda_0$ is related to microscopic parameters in QCD; see ahead to Eq.~(\ref{Lambda0}).)
Hence, in order to make up the observed dark matter of the universe, the axion mass is normally expected to be around $m_a\sim 10^{-5}$\,eV. 

However, there are reasons to consider axions that are either somewhat heavier or lighter than this value. In particular, there is some motivation to consider heavier, say $\m\sim 0.1 - 1$\,eV, to explain the anomalous cooling in horizontal branch stars (the so-called ``HB hint" \cite{Giannotti2016}). Also, it is of interest to consider axions that are lighter, say $\m\sim 10^{-10} - 10^{-9}$\,eV, to correspond to 
$\f$ closer to scales suggested by fundamental physics, such as in grand unified theories (GUT) or string theory \cite{Svrcek2006}.
In canonical axion models, these lighter or heavier axions cannot ordinarily give rise to the observed dark matter abundance (for lighter axions, one may appeal to selection effects in the preinflationary scenario \cite{Pi1984,Linde1991,Wilczek2004,Tegmark2006,Fox2004,Hertzberg2008}, though this relies on several assumptions).

In this work, we propose a simple mechanism that can alter the axion abundance from the above value. 
We are motivated by the fact that axion models appeal to extremely high energy physics associated with PQ breaking $\sim \f$, as yet untested. The details of the physics governing this may involve richer dynamics than assumed in the simplest models. To capture possible new behavior, here we introduce a second light scalar that couples to the complex Peccei-Quinn field, in a way that renders the effective symmetry-breaking scale dynamical.  This, as we show, leads to an alteration in the misalignment production of axions which affects the relic abundance.\footnote{Very interesting previous work has been done involving axion dynamics different from those discussed in our work, also seeking to alter the abundances and expand the window of viable axion parameters. See, for example, Ref.s~\cite{Dvali1995,Heurtier2021,Co2019,Co2019a,Co2018,Co2020,Kitano2021,Nakayama2021,Kobayashi2021,Dienes2017,Dienes2016,DiLuzio2020}. In particular, Ref.~\cite{Dvali1995} discusses adjustments which are controlled by the vacuum expectation value of other scalars, though they focus on modifying gauge couplings in this way. }
We find that the dynamics are surprisingly rich, and can lead to either an enhanced or lower abundance, depending on parameters. We provide simple analytical estimates for the final abundance for both the axion and the new scalar, and we accompany this with detailed numerical results; albeit within the simple homogeneous field approximation. 

The outline of our paper is as follows: 
In Section \ref{Std} we begin by recaping the standard axion models and their abundance.
In Section \ref{Model} we introduce our new model and provide simple analytical estimates for the correspondence abundance.
In Section \ref{Numerical} we perform a numerical analysis of the classical field equations.
In Section \ref{Constraints} we discuss constraints from particle physics.
Finally, in Section \ref{Conclusions} we conclude.

\section{Standard PQ Scale Mechanism}\label{Std}

In canonical models of the axion, a complex PQ field $\Phi=\rho\,e^{i\theta}$ is introduced, which carries a new global symmetry. The angular part $\theta$ is a (pseudo-)Goldstone boson that emerges when the symmetry is spontaneously broken, which is the axion. The standard Lagrangian for $\Phi$ is given by
\beq
\mathcal{L} = \sqrt{-g}\left[{1\over2}|\partial\Phi|^2-{\lambda\over 4}(|\Phi|^2-\f^2)^2-V(\theta,T)\right]
\label{LStandard}\eeq
Here the effective potential for the axion $V$ is temperature dependent, as it takes into account the effects from interactions with Standard Model particles in thermal equilibrium. It is given approximately by
\beq
V(\theta,T)=\Lambda(T)^4(1-\cos\theta)
\eeq
The cosine factor is not precise, as it is only true in the dilute instanton approximation, while a more accurate treatment gives a moderately altered function of $\theta$ \cite{DiCortona2016}; the details will not be central in this work.  The temperature dependent scale is roughly given by
\bea
\Lambda(T)^4=\Bigg{\{}
\begin{array}{l}
\Lambda_0^4(\Tqcd/T)^8,\,\,\,T\gg\Tqcd\\
\Lambda_0^4,\,\,\,\,\,\,\,\,\,\,\,\,\,\,\,\,\,\,\,\,\,\,\,\,\,\,\,\,\,T\ll\Tqcd
\end{array}
\eea
where the temperature of the QCD phase transition is of order of the QCD scale $\Tqcd\sim 200$\,MeV.
The power of 8 is not necessarily precise, as indicated by some lattice simulations, but it will suffice for our purposes (see Refs.~\cite{Bazavov2012,DiLuzio2020} and references therein for more information).
The low temperature value is given by
\beq
\Lambda_0^2={\sqrt{m_u m_d}\over m_u+m_d}\,f_\pi\,m_\pi
\label{Lambda0}\eeq
which evaluates to $\Lambda_0\approx 90$\,MeV. 

\subsection{Standard Axion Evolution}

At low energies, the radial field $\rho$ gets frozen-in at its minimum energy value $\rho=\f$. This leaves the low energy Lagrangian for just the angular field (axion) $\theta$ as
\beq
\mathcal{L} = \sqrt{-g}\left[{1\over2}\f^2(\partial\theta)^2-V(\theta,T)\right]
\eeq
For an FRW universe, in co-moving coordinates, the metric is
$g_{\mu\nu}=\mbox{diag}(1,-a^2,-a^2,-a^2)$.
 The prefactor in the Lagrangian is
$\sqrt{-g}=a^3$. 

The classical field equation of motion for $\theta$ (ignoring metric perturbations) is then
\beq
\ddot\theta+3H\dot\theta-{\nabla^2\theta\over a^2}+{\Lambda(T)^4\over\f^2}\sin\theta=0
\eeq
If we ignore the spatial variation in $\theta$, and for small angles we approximate $\sin\theta\approx\theta$, then this simplifies to
\beq
\ddot\theta+3H\dot\theta+{\Lambda(T)^4\over\f^2}\theta=0
\eeq
We also need to know how the Hubble rate $H=\dot{a}/a$ changes in time. Let us recap this very well-known behavior, as follows. It is related
to the energy density of the universe $\rho$ by
$H^2 = \rho/(3\mpl^2)$, 
where $\mpl=1/\sqrt{8\pi G}\approx 2.4\times 10^{18}$\,GeV is the (reduced) Planck mass. 
In a radiation dominated universe (roughly the first 70,000 years) the energy density is given in terms of temperature by
\beq
\rho ={\pi^2\over30}g\, T^4
\eeq
where $g$ is the number of active relativistic degrees of freedom. Since temperature redshifts as $T\propto 1/a$, this leads to Hubble changing with time as
\beq 
H={1\over 2\,t}, 
\eeq
as is very well known.

\subsection{Standard Relic Abundance}

Now let us summarize a simple way to estimate the axion abundance in standard models through the misalignment mechanism.
Firstly, the axion is friction dominated and almost frozen in its potential at some initial angle $\theta_i$. Then it
starts to roll at a temperature $\Ts$ when 
\beq
3H(\Ts)\approx m(\Ts)=\Lambda(\Ts)^2/\f
\eeq
Using the above expressions for $H$ and $\Lambda$, we can solve for $\Ts$ to obtain
\beq
\Ts = \left(\sqrt{10}\,\mpl\Tqcd^4\Lambda_0^2\over\f\sqrt{g_*}\,\pi\right)^{1/6}
\eeq
The corresponding number density of axions at this time is
\beq
n(\Ts)={\rho_a(\Ts)\over m(\Ts)}
\eeq
where the energy density of axions is
\beq
\rho_a(\Ts)\approx{1\over2}\Lambda(\Ts)^4\langle\theta_i^2\rangle
\eeq
Here we spatially average over the initial $\theta_i$. One can correct this formula accounting for anharmonicity of the potential, but this form is acceptable so long as $\theta_i$ is not near $\pm\pi$.   

These axions have their number density redshift in a simple way, since they are conserved. The final energy density then just requires us to multiply by the final (zero temperature) axion mass $\m$. A useful way to represent this is through the ratio of axion energy density to photon number density at late times \cite{Hertzberg2008}
\beq
\xi_a\equiv{\rho_a(T_0)\over n_\gamma(T_0)} 
\eeq
This can be written out as
\beq
\xi_a = {\m\over m(\Ts)}{\rho_a(\Ts)\over n_\gamma(T_0)}{s(T_0)\over s(\Ts)}
\eeq
where $s(T)$ is the total entropy density and $n_\gamma(T)$ is the photon number density
\beq
s(T)={2\pi^2\over45}g_{s}\,T^3,\,\,\,\,\,\,n_\gamma(T)={2\zeta(3)\over\pi^2}T^3
\eeq
This takes into account the fact that there are changes in the number of species through annihilations. If one ignores this detail, this equation simply says $\rho_a(T_0)\sim(\m/m(\Ts))(T/\Ts)^3\rho_a(\Ts)$, which is somewhat intuitive.

Putting this together, we obtain
\beq
\xi_{a,std}\approx 
2{g_*^{7/12}g_{s0}\over g_{s*}}
{\Lambda_0^{5/3}\f^{7/6}\over \Tqcd^{2/3}\mpl^{7/6}}
\langle\theta_i^2\rangle
\label{xistandard}\eeq
This shows the $\xi\propto\f^{7/6}$ dependence on the PQ scale, summarized earlier in Eq.~(\ref{Omaxion}). Note that this ratio $\xi_a=\rho_a/n_\gamma$ is useful because it is time independent, since at late times the numerator and denominator redshift in the same way. The observed value of $\xi$ for cold dark matter is 
\beq
\xi_{obs}\approx 2.9\,\mbox{eV}
\eeq
which (not coincidentally) is on the order of the temperature of the universe at equality. Inserting parameters, one finds that the observed abundance is realized by Eq.~(\ref{xistandard}) for $\f$ of a few times $10^{11}$\,GeV, or an axion mass $m_a$ around $10^{-5}$\,eV, as mentioned earlier.

\section{Dynamical PQ Scale Mechanism}\label{Model}

We now formulate a two-field model, involving the complex PQ field $\Phi$ and a new light scalar field $\chi$. The idea is to modify the dynamics of the PQ radial field $\rho=|\Phi|$ so that it evolves in an interesting way. This means that the effective Peccei-Quinn (PQ) scale, and indeed the effective mass of the axion, will also evolve. As we shall see, this will lead to alterations in the relic abundance of axions for a given (late-time) axion mass.

We update the standard Lagrangian of Eq.~(\ref{LStandard}) to the following 
\bea
\mathcal{L} = \sqrt{-g}\Big{[}{1\over2}|\partial\Phi|^2-{\lambda\over 4}(|\Phi|^2-f(\chi)^2)^2-V(\theta,T)\nonumber\\
+{1\over2}(\partial\chi)^2-{1\over2}\mchi^2\,\chi^2\Big{]}
\label{LNew}\eea
Here, we have promoted the PQ scale $\f$ to now become effectively dynamical, as it is controlled by the value of the second field $\chi$ through the function $f(\chi)$. We wish to consider situations in which the $\chi$ field evolves such that $f(\chi)$ relaxes to fixed values,  $\f$, at late times, but can be different at early times, either larger or smaller. We can parametrize this as 
\beq
f(\chi)^2=\f^2\,\mudim(\chi)
\eeq
where $\mudim$ is a dimensionless function. Some concrete choices are
\beq
\mudim(\chi)=\left(1+\chi^2/\Sc^2\right)^{\pm 1}\label{alpha}
\eeq
where $\Sc$ is a new scale that controls the size of these effects. 
We will consider both positive, $+1$, and negative, $-1$, exponents in this work. For positive (negative) exponents, the $\alpha$ function ensures that for nonzero $\chi_i$ the value of $f$ in the early universe is relatively large (small), before {\em decreasing} ({\em increasing}) toward $f_a$ in the later universe when $\chi$ itself relaxes toward zero. We will refer to these as the decreasing-PQ-scale (DPQ) model and increasing-PQ-scale (IPQ) model, respectively.
Note that we have chosen the potential for $\chi$ to be just quadratic ${1\over2}\mchi^2\,\chi^2$ for simplicity, though we could consider more complicated potentials. The Lagrangian for $\chi$ is endowed with a $\mathbb{Z}_2$ symmetry $\chi\to-\chi$ for simplicity. 
As we shall see, we will need the $\chi$ field to be light for new interesting behavior. Its lightness does not seem obvious to justify since it does not carry a shift symmetry. If one estimates loop corrections to the mass with a UV cutoff, then typically parameters need a cancellation between the bare mass and the loop correction for the renormalized mass to be small (although one would not see this using dimensional regularization). 
On the other hand, one may explore more general models, in which the $\chi$ mass is protected by extending the $U(1)$ PQ symmetry to a larger symmetry group.
In any case, we leave analysis of microscopic considerations for future work, and focus on the phenomenology here in this paper as a starting point. Under cosmic evolution, one expects $\chi\to0$ at late times, ensuring $f\to\f$ at late times too.

\subsection{Non-Standard Axion Evolution}

As before, we can still assume that the PQ field is extremely heavy, so it once again gets stuck at the value that minimizes its potential in the very early universe. In this case, that is
\beq
\rho\to f(\chi)
\eeq
so its value is controlled by the dynamics of the new light field $\chi$.
This leaves a low energy Lagrangian for the angular field (axion) $\theta$ and $\chi$ as
\beq
\mathcal{L} = \sqrt{-g}\left[{1\over2}f(\chi)^2(\partial\theta)^2-V(\theta,T)+{1\over2}(\partial\chi)^2-{1\over2}\mchi^2\,\chi^2\right]
\eeq
In principle, one should also add the kinetic term for $\rho$, using ${1\over2}(\partial\rho)^2\to{1\over2}f'(\chi)^2(\partial\chi)^2$. However in the regime $S\gg f_a$, these corrections are small, and even for $S\sim f_a$, the corrections are only moderate. We shall focus on the regime $S\gtrsim f_a$ in this paper and therefore ignore these corrections for brevity. We have checked that our results are not significantly affected by these corrections.

The corresponding classical equations of motion for the fields are
\bea
&&\ddot\theta+\left(3H+2{f'(\chi)\over f(\chi)}\dot\chi\right)\!\dot\theta-{\nabla^2\theta\over a^2}+{\Lambda(T)^4\over f(\chi)^2}\sin\theta=0\,\,\,\,\\
&&\ddot\chi+3H\dot\chi-{\nabla^2\chi\over a^2}+\mchi^2\,\chi-f'(\chi)f(\chi)\dot\theta^2 = 0
\eea
(plus one can readily include the corrections from the kinetic term of $\rho$, if desired).

It is useful to recast these equations in terms of dimensionless variables. We define a dimensionless time variable $\tau$, a dimensionless second field $\Y$, a dimensionless coupling $\Fc$, a dimensionless temperature dependent scale $\LamDim$, and the ratio of masses $\mratio$ as follows
\beq
\tau\equiv\m\,t,\,\,\,\,\Y\equiv{\chi\over S},\,\,\,\,\Fc\equiv{\f\over S},\,\,\,\,\LamDim(T)\equiv{\Lambda(T)^4\over\Lambda_0^4},\,\,\,\,\mratio\equiv{\mchi\over\m}
\eeq
We also once again ignore spatial variations, assume small angles $\sin\theta\approx\theta$, and use $H=1/(2\,t)$ in the radiation era to obtain the dimensionless equations of motion
\bea
&&\theta_{\tau\tau}+\left({3\over2\,\tau}+{\mudim'(\Y)\over \mudim(\Y)}\Y_\tau\right)\!\theta_\tau+{\LamDim(T)\over\mudim(Y)}\theta=0\,\,\,\,\\
&&\Y_{\tau\tau}+{3\over2\,\tau}\Y_\tau+\mratio^2\Y-{\Fc^2\over2}\mudim'(Y)\theta_\tau^2 =0
\eea
Note that we can rewrite the temperature dependence in $\LamDim$ in terms of the time variable instead. This becomes
\bea
\LamDim(\tau)=\Bigg{\{}
\begin{array}{l}
(\tau/\tauqcd)^4,\,\,\,\tau\ll\tauqcd\\
1,\,\,\,\,\,\,\,\,\,\,\,\,\,\,\,\,\,\,\,\,\,\,\,\,\,\tau\gg\tauqcd
\end{array}
\eea
where the dimensionless time scale of the QCD phase transition is found to be
\beq
\tauqcd={\sqrt{90}\,\Lambda_0^2\over2\pi\sqrt{\gqcd}\,\Tqcd^2}{\mpl\over\f}
\eeq

\subsection{Simple Estimates for Altered Abundance}\label{sec:estimates}

Suppose the $\chi$ field is initially at the value $\chi_i$; the corresponding dimensionless initial value is $\Y_i=\chi_i/S$. At early times the $\chi$ field will be effectively frozen at this value with an effective PQ scale of
\beq
f_i=\f\sqrt{\mudim(\Y_i)}
\eeq
Assuming the coupling between the two fields is not too significant, one anticipates that the number density of axions is given by the same formula as before (\ref{xistandard}), except with this effective PQ-scale $f_i$ instead. So an estimate is
\beq
{\xi_{a}\over \m} = {1\over m(\Ts(f_i))}{\rho_a(\Ts(f_i))\over n_\gamma(T_0)}{s(T_0)\over s(\Ts(f_i))}
\eeq
Using Eq.~(\ref{xistandard}), but with $\f\to f_i$ and then multiplying throughout by $\m$ (the late-time value), we obtain the alteration factor of
\beq
{\xi_{a}\over\xi_{a,std}}=\left(f_i\over\f\right)^{13/6}=\mudim(\Y_i)^{13/12}
\label{xiax}\eeq

We would also like an estimate of the $\chi$ abundance. Again, we shall do this assuming the coupling between the fields is not too significant. Just like for the axion, the $\chi$ field is frozen in its potential until a temperature $\Tchi$ of
\beq
3H(\Tchi)\approx \mchi
\eeq
Note that in this model, we assume the $\chi$ field has a temperature independent mass, and so this relation is relatively simple. Solving for $\Tchi$, we find
\beq
\Tchi = \left(\sqrt{10}\,\mpl\mchi\over\sqrt{g_\chi}\,\pi\right)^{1/2}
\eeq
The corresponding energy density of $\chi$ particles at this time is
\beq
\rho_\chi(\Tchi)\approx{1\over2}\mchi^2\langle\chi_i^2\rangle
\eeq
The late-time ratio of $\chi$ energy density to photon number density is 
\beq
\xi_\chi = {\rho_\chi(\Tchi)\over n_\gamma(T_0)}{s(T_0)\over s(\Ts)}
\eeq
This gives
\beq
\xi_\chi\approx 
2{g_\chi^{3/4}g_{s0}\over g_{s\chi}}
{\sqrt{\mchi}\over \mpl^{3/2}}\langle\chi_i^2\rangle
\label{chistandard}\eeq
In terms of the above dimensionless parameters, we can express this in terms of the standard axion abundance as
\beq
{\xi_\chi\over\xi_{a,std}}\approx {g_\chi^{3/4}g_{s*}\,\Tqcd^{2/3}\over g_*^{7/12}g_{s\chi}\,\Lambda_0^{2/3}}
{\f^{1/3}\over\mpl^{1/3}}
{\sqrt{\mratio}\,\langle Y_i^2\rangle\over\Fc^2\langle\theta_i^2\rangle}
\label{xichi}
\eeq

Comparing Eq.~\ref{xichi} and Eq.~\ref{xiax}, we can see that both fields have a relic abundance related to $Y_i$. Thus it is also interesting to consider the ratio
\beq \frac{\xi_\chi}{\xi_{a}}\approx{g_\chi^{3/4}g_{s*}\,\Tqcd^{2/3}\over g_*^{7/12}g_{s\chi}\,\Lambda_0^{2/3}}
{\f^{1/3}\over\mpl^{1/3}}
{\sqrt{\mratio}\,\langle Y_i^2\rangle\over\Fc^2\langle\theta_i^2\rangle}\mudim(\Y_i)^{-13/12}
\label{xichixia}
\eeq
where $\xi_a$ is the axion abundance in the dynamical PQ model. If the resultant cosmology involves the axion as the dark matter, we demand that the $\chi$ field has a negligible relic abundance, and thus that this ratio $\xi_\chi/\xi_{a}\ll 1$. However, the only strict requirement is that both the axion and the new field do not overclose the universe.

\section{Numerical Analysis}\label{Numerical}

We will discuss separately the two theory choices, that of either decreasing or increasing the PQ scale, by discussing the two particular models with $\alpha(\chi)=(1+\chi^2/S^2)$ (DPQ model) and $\alpha(\chi)=(1+\chi^2/S^2)^{-1}$ (IPQ model). 
First, we will illustrate the general dynamics in sections~\ref{sec:DPQ} and~\ref{sec:IPQ}, showing that enhancement and reduction of abundance is possible, as expected by the above simple analysis. In these sections, we will not, however, present results with extreme changes in axion masses with the resulting desired abundance, as this requires extreme changes in parameters, which is difficult to efficiently handle numerically. Second, to complement this, in section~\ref{sec:paramdep} we will give evidence that our analytical theory is fairly accurate by examining a broad range of parameter space and finding good agreement; the drawback is that the parameter choices will not always be of the most interest.

\subsection{Analysis of Decreasing-PQ-Scale Model}\label{sec:DPQ}

\begin{figure}[t]
\centering
\includegraphics[width=\linewidth]{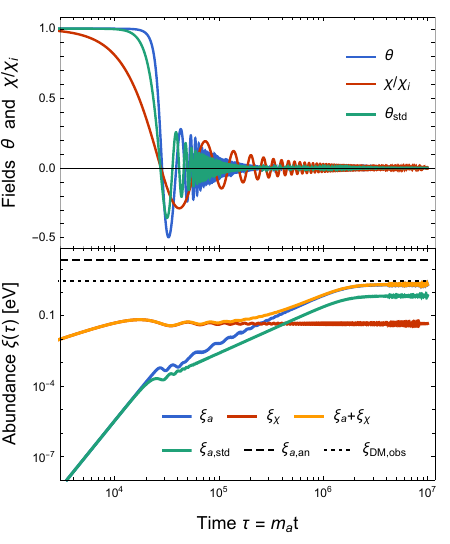}
\caption{A realization of the DPQ model with $m_a = 10^{-4} \mbox{ eV}$, $m_\chi=10^{-8} \mbox{ eV}$, $F=0.1$, and $Y_i=\chi_i/S=5$. The top panel shows the evolution of the axion ($\theta$) and $\chi$ field compared to the evolution of the standard QCD axion ($\theta_{\text{std}}$). The lower panel shows the abundance parameter $\xi\equiv \rho/n_\gamma$ for the axion ($\xi_a$), the $\chi$ field ($\xi_\chi$), their sum, and the standard axion ($\xi_{a,\text{std}}$). The dashed line indicates the analytical prediction $\xi_{a,an}$ from Eq.~\ref{xiax}, and the dotted line shows $\xi_{\text{DM,obs}}\approx 2.9 \mbox{ eV}$, the observed DM abundance. This axion mass can, in particular, be interesting for experimental searches targeting this mass, such as MADMAX. In this instance, a heavier axion is adapted to reproduce the observed DM abundance; the $\chi$ field is not light enough to begin oscillations after the axion, rendering the analytical prediction (dashed line) inaccurate.}
\label{fig:DPQ1}
\end{figure}

Let us first consider the DPQ model, which can account for dark matter axions with larger masses than usual. This choice has the feature that the resulting relic abundance of the axion is greater than the typical misalignment product of an axion of the same mass, due to the fact that the effective Peccei-Quinn scale is initially larger than its late-time value. Let us consider, for example, a QCD axion of mass $10^{-4} \mbox{ eV}$. The misalignment mechanism, with generic choices for initial conditions (i.e. $\theta_i\sim1$), predicts that the relic abundance of this axion would be about an order of magnitude smaller than the observed abundance of dark matter. In the DPQ model, the same axion can reproduce the observed relic abundance of dark matter, depending on parameters. The full dynamics of the fields is presented in the upper panel of Fig.~\ref{fig:DPQ1}, while the lower panel shows the corresponding energy densities.

The realization of the DPQ model in fig.~\ref{fig:DPQ1} can be of interest when considering some experimental searches for axion dark matter. For instance, the MADMAX experiment \cite{Brun2019} targets axion masses around $10^{-4}$ eV; if such an axion is found to be as abundant as dark matter, or at least more abundant than predicted by the standard misalignment mechanism, this abundance could be explained by a dynamical PQ scale model like the DPQ model.

\begin{figure}[t]
\centering
\includegraphics[width=\linewidth]{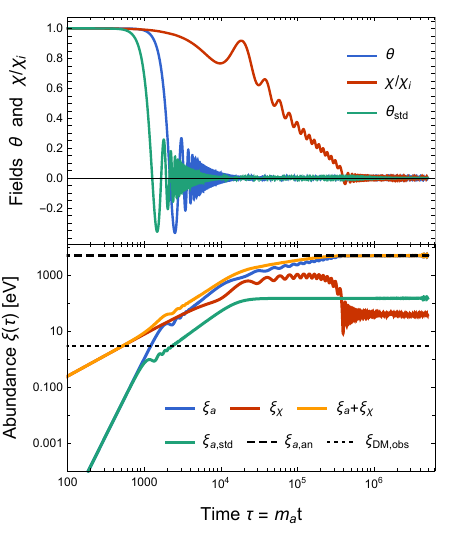}
\caption{A realization of the DPQ model with $m_a = 10^{-6} \mbox{ eV}$, $m_\chi=2\times 10^{-10} \mbox{ eV}$, $F=0.1$, and $Y_i=\chi_i/S=5$. The top panel shows the evolution of the axion ($\theta$) and $\chi$ field compared to the evolution of the standard QCD axion ($\theta_{\text{std}}$). The lower panel shows the abundance parameter $\xi\equiv \rho/n_\gamma$ for the axion ($\xi_a$), the $\chi$ field ($\xi_\chi$), their sum, and the standard axion ($\xi_{a,\text{std}}$). The dashed line indicates the analytical prediction $\xi_{a,an}$ from Eq.~\ref{xiax}, and the dotted line shows $\xi_{\text{DM,obs}}\approx 2.9 \mbox{ eV}$, the observed DM abundance. In this instance, a lighter axion is used to illustrate that the analytical estimate can accurately predict the axion abundance in the DPQ model.}
\label{fig:DPQ2}
\end{figure}

Note that, in this case, although the $\chi$ field is four orders of magnitude lighter than the axion, it nonetheless begins its oscillations before the axion does because of the axion's effectively smaller mass at early times (the axion's mass is temperature dependent in the early universe). For this reason, as seen in Fig.~\ref{fig:DPQ1}, the increased abundance of the axion (dark blue) compared to a standard axion (green) is milder than the analytical formulas predict (dashed black line), since these formulas assume that the $\chi$ field begins oscillations later than the axion does. Figure~\ref{fig:DPQ2} shows a realization of the model with $m_a=10^{-6}\mbox{ eV}$, $m_\chi=2\times10^{-10} \mbox{ eV}$, $Y_i=5$, and $F=0.1$, demonstrating a case where the $\chi$ field is light enough to begin its oscillations later than the axion, and thus the enhanced axion abundance is more dramatic.

It would be interesting, now, to consider axions even heavier than those shown in Fig.s~\ref{fig:DPQ1} and~\ref{fig:DPQ2} which have some observational motivations. For example, for a particular axion model (i.e. KSVZ, DFSZ, and further variations), the value of the Peccei-Quinn scale (or equivalently the axion's mass) predicts a coupling strength for an axion-photon interaction of the form $\phi F_{\mu\nu}\tilde{F}^{\mu\nu}$. The observed anomalous cooling in horizontal branch (HB) stars could be explained by a cooling mechanism in which photons convert to axions and carry away some energy from the star. This mechanism could account for the observed anomalous cooling if the axion mass were on the order of $m_a \sim 0.1 - 1 \mbox{ eV}$ \cite{Giannotti2016}. In addition, this mass regime may be of interest for experiments such as LAMPOST~\cite{Chiles2022}, which searches for axions with masses $0.1 - 10$ eV. In the standard misalignment mechanism, these axions would have a very small relic abundance, $\lesssim 10^{-6}$ times the observed dark matter abundance, much too small to be significant. 

However, in the case of an axion coupled to another scalar as in our dynamical PQ scale mechanism, this heavier axion can be produced in appreciable quantities. For example, from Eq.~\ref{xiax}, we can deduce that with the initial field value $Y_i\approx 280$, an axion of mass $m_a \sim 1 \mbox{ eV}$ can reproduce the observed dark matter abundance. Crucially, in this mass regime, the axion begins oscillations before its QCD effect-induced mass reaches its zero-temperature value, so we must have $m_\chi < 10^{-7}\, m_a$ in order for the $\chi$ field to begin oscillations later than the axion\footnote{In principle, it is not necessary for the axion to begin oscillations first, and a viable cosmology could result from a slightly heavier $\chi$. However, in this analysis, we rely on the analytical estimates in section~\ref{sec:estimates}, which assume the $\chi$ field begins oscillations later, and thus we do not make claims here for the opposite case.}. With a choice of the coupling parameter of $F=0.1$, we can further deduce from Eq.~\ref{xichixia} that the $\chi$ field has a relic abundance which is heavily suppressed, $\sim 10^{-6}$ times smaller than the (dark matter) axions. This set of parameters therefore results in a late-time cosmology which is very similar to a universe with cold dark matter (CDM) or standard axion dark matter. The important difference is that this axion has a mass which, without the dynamical PQ mechanism, would not produce a sufficiently high relic abundance to account for DM.

So far, we have deduced the choices of parameters using the analytical formulas developed in section~\ref{sec:estimates}; see section~\ref{sec:paramdep} for a numerical analysis of these formulas. A direct verification of the dynamics of a $m_a\sim 1 \mbox{ eV}$ axion presents numerical difficulties, in part because it requires integrating through a large amount of time (from before the QCD phase transition to after the axion oscillations settle into matter-like behavior); we therefore leave this verification for future work which would require more sophisticated numerical methods.

\begin{figure}[t]
\centering
\includegraphics[width=\linewidth]{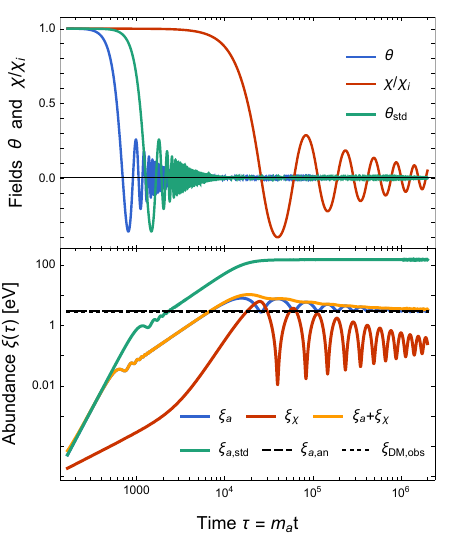}
\caption{A realization of the IPQ model with $m_a = 10^{-6} \mbox{ eV}$, $m_\chi=5\times 10^{-12} \mbox{ eV}$, $F=0.5$, and $Y_i=\chi_i/S=6$. The top panel shows the evolution of the axion ($\theta$) and $\chi$ field compared to the evolution of the standard QCD axion ($\theta_{\text{std}}$). The lower panel shows the abundance parameter $\xi\equiv \rho/n_\gamma$ for the axion ($\xi_a$), the $\chi$ field ($\xi_\chi$), their sum, and the standard axion ($\xi_{a,\text{std}}$). The dashed line indicates the analytical prediction $\xi_{a,an}$ from Eq.~\ref{xiax}, and the dotted line shows $\xi_{\text{DM,obs}}\approx 2.9 \mbox{ eV}$, the observed DM abundance. This axion mass may be of interest when considering axion searches such as ADMX. Note that the analytical prediction (dashed), observed DM (dotted), and $\xi_a$ (blue) curves all coincide, displaying the success of the IPQ model in adapting this particular axion to a viable cosmological outcome.}
\label{fig:IPQ3}
\end{figure}

\begin{figure}[t]
\centering
\includegraphics[width=\linewidth]{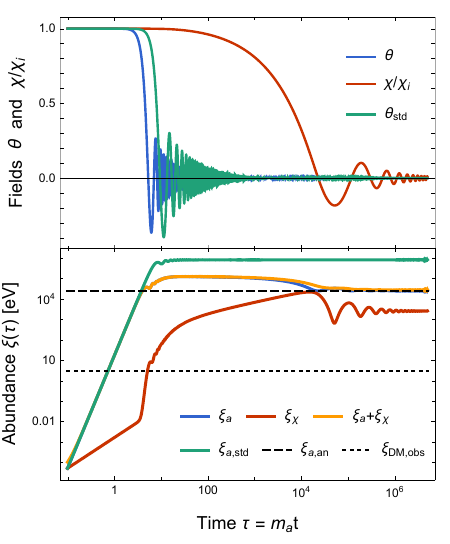}
\caption{A realization of the IPQ model with $m_a = 6\times10^{-10} \mbox{ eV}$, $m_\chi=6\times 10^{-15} \mbox{ eV}$, $F=0.5$, and $Y_i=\chi_i/S=5$. The top panel shows the evolution of the axion ($\theta$) and $\chi$ field compared to the evolution of the standard QCD axion ($\theta_{\text{std}}$). The lower panel shows the abundance parameter $\xi\equiv \rho/n_\gamma$ for the axion ($\xi_a$), the $\chi$ field ($\xi_\chi$), their sum, and the standard axion ($\xi_{a,\text{std}}$). The dashed line indicates the analytical prediction $\xi_{a,an}$ from Eq.~\ref{xiax}, and the dotted line shows $\xi_{\text{DM,obs}}\approx 2.9 \mbox{ eV}$, the observed DM abundance. In this instance, the IPQ model can suppress the abundance of a lighter axion with $f_a$ motivated by the GUT scale, though not down to the observed DM abundance.}
\label{fig:IPQ2}
\end{figure}

\subsection{Analysis of Increasing-PQ-Scale Model}\label{sec:IPQ}

Now let us examine the IPQ model.
Here the relic abundance of the axion is suppressed compared to its usual misalignment value. 

In standard models, if one considers a rather large $f_a\gg 10^{12}$\,GeV PQ-scale the standard misalignment abundance of such an axion would be much too large (unless one appeals to very small $\theta_i$, which is not our focus). For instance, if we consider a slightly larger PQ scale ($f_a\sim 6\times 10^{12}$\,GeV), such that the axion mass is $m_a=10^{-6}$\,eV, the standard misalignment mechanism predicts a relic abundance that would overclose the universe. Instead, in the IPQ model with $m_\chi=5\times10^{-12}$\,eV, $F=0.5$, and $Y_i = \chi_i/S=6$, the abundance can be suppressed down to the observed DM abundance, preventing the overclosing of the universe. This realization is shown in Fig.~\ref{fig:IPQ3}, with the evolution of the fields in the top panel and the abundances in the bottom panel.

This realization of the IPQ model may be relevant when considering some axion search experiments, in particular ADMX~\cite{Bartram2021} which has focused on an axion mass range in the vicinity of $10^{-6}$ eV. In the last years, ADMX has been able to heavily constrain many models of axion dark matter near this mass, but modifications of these models through a dynamical PQ scale may change this conclusion. Further, an axion of this mass could make up a fraction of the dark matter, which would lessen the constraint from ADMX and could be realized in a model like the IPQ model.

As another important example, one might be motivated to consider values for the PQ scale approaching the GUT scale e.g. $f_a\sim 10^{16} \mbox{ GeV}$, corresponding to  $m_a\sim 6\times 10^{-10}\mbox{ eV}$.
As shown in the bottom panel of Fig.~\ref{fig:IPQ2}, the IPQ model allows for a smaller abundance of this axion. In this iteration of the model, the $\chi$ field's mass is $m_\chi = 6\times 10^{-15}\mbox{ eV}$, the coupling parameter is $F=0.5$, and the initial field value is $Y_i=5$.

This scenario, however, still leads to an overproduction of both the axion and the $\chi$ field to the point of overclosing the universe. In order to avoid this, the parameters of the model must be chosen such that the suppression of the axion abundance is very significant. 
The desired axion abundance can be achieved nearly trivially by a choice of the initial field value $Y_i\approx 320$. However, as indicated in Eq.~\ref{xichi}, and shown later in Fig.~\ref{fig:xivY2}, increasing $Y_i$ amplifies the abundance of the $\chi$ field, and the energy density of the universe is still too large.

Let us consider, for the moment, an energy scale that is a bit lower but still desirable, $f_a\sim 10^{15} \mbox{ GeV}$, corresponding to roughly  $m_a\sim 5\times 10^{-9}\mbox{ eV}$. With this choice of late-time PQ scale and axion mass, one can suppress the abundance of the axion down to the DM abundance by choosing $Y_i=108$. In order to avoid overproduction of the $\chi$, the remaining model parameters ($F$, the effective strength of the coupling of the two fields, and $m_\chi$) should be modified. It should be noted that, as a simplifying assumption, our analysis has taken the coupling between the fields to be relatively weak ($F\ll 1$). Thus, taking $F=0.5$ as a maximum, we can modify $m_\chi$ accordingly (smaller $F$ requires an even lighter $m_\chi$). In order for the $\chi$ to be subdominant\footnote{We require $\xi_\chi$ to be $< 5\%$ of the DM. This is to avoid CMB constraints on ultralight fields of this mass. See, e.g. \cite{HloZek2017}.}, a mass $m_\chi\sim 10^{-28}$\,eV is needed. Note that this corresponds to the onset of oscillations being right around the epoch of matter-radiation equality. This is perhaps the lightest mass for which our analysis can be valid.

Interestingly, such a light field may have implications for the tension in large-scale structure measurements between cosmic shear surveys and CMB inferences, known as the $S_8$ tension (see \cite{Asgari2020,Joudaki2020,Heymans2013,Hildebrandt2020,Abbott2018,Hikage2019,Heymans2021}). Such light fields are known to suppress structure growth, and although they are heavily constrained, it has been shown that even in small abundances they may have noticeable effects on the matter power spectrum of the universe (see \cite{Allali2021} and also \cite{Fung2021,Luu2021}, among others).

If, instead, one were insistent on $f_a\sim 10^{16}\mbox{ GeV}$, then in order to produce a viable cosmology with $Y_i\approx 320$ and $F=0.5$, the $\chi$ must be so light that it indeed behaves as dark energy today, $m_\chi\sim 10^{-33} \mbox{ eV}$, and thus we may not confidently rely on the present analysis.



Once more, we make these predictions using the analytical formulas developed in section~\ref{sec:estimates}; see section~\ref{sec:paramdep} for a numerical analysis of these formulas. Similarly, here, the numerical difficulties arise from integrating for long times, but now specifically it is the onset of $\chi$ oscillations which begin quite late (compared to the time scale for the onset of axion oscillations). Altogether, while we can be confident that somewhat larger $f_a$ than normal can be accommodated in our framework, we do not have concrete numerical evidence that $f_a$ much, much higher, say GUT scale, can be accommodated. We leave a more sophisticated numerical analysis of the relevant parameters, as well as variations on our setup, for future work.

\subsection{Parameter Dependence of Abundance}\label{sec:paramdep}

The analytical formulas that predict the relic abundances of the fields in the dynamical PQ mechanism, in particular Eq.~\ref{xichixia}, are useful for determining what model parameter choices result in viable cosmologies. For example, certain choices lead to overproduction of either the axion or the $\chi$ field, leading to energy densities which are inconsistent with observations and would overclose the universe. We present here a numerical analysis of varying the initial field value $Y_i=\chi_i/S$, the mass of the $\chi$ field $m_\chi$, and the coupling parameter $F$.

\begin{figure}[t]
\centering
\includegraphics[width=\linewidth]{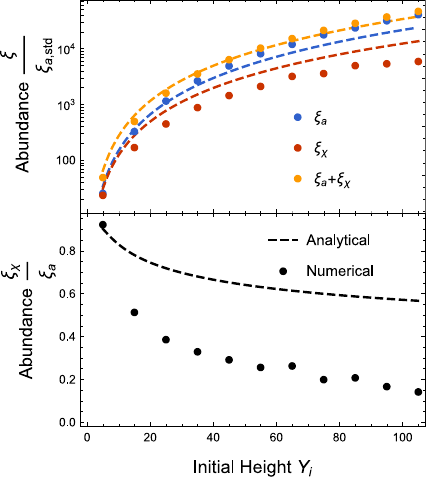}
\caption{The ratio of various abundances in the DPQ model ($\alpha(\chi)=(1+\chi^2/S^2)$) to the abundance of axions in the standard theory, $\xi/\xi_{a,std}$, is plotted for various initial field values $Y_i=\chi_i/S$, showing the enhancement in axion and $\chi$ abundance resulting from tuning this parameter. The ratio $\xi_\chi/\xi_a$ in the DPQ model is also shown. The axion mass is fixed at $m_a=10^{-10} \mbox{ eV}$, the $\chi$ mass is $m_\chi = 5\times 10^{-4}\, m_a$, and the coupling parameter is $F=0.1$. The dashed lines show the analytical estimates, while the points correspond to individual numerical solutions to the field equations.}
\label{fig:xivY1}
\end{figure}

\begin{figure}[t]
\centering
\includegraphics[width=\linewidth]{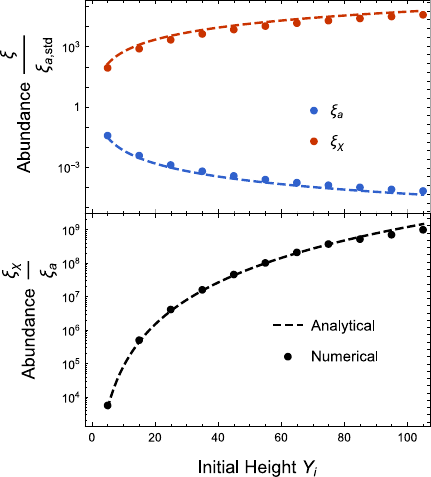}
\caption{The ratios of abundances in the IPQ model ($\alpha(\chi)=(1+\chi^2/S^2)^{-1}$) are plotted for various initial field values $Y_i=\chi_i/S$, showing the suppression of axion abundance resulting from tuning this parameter. In the upper panel, the sum $\xi_a+\xi_\chi$ is not shown since it is nearly identical to $\xi_\chi$ as in this case the $\chi$ field is dominant. The axion mass is fixed at $m_a=10^{-10} \mbox{ eV}$, the $\chi$ mass is $m_\chi = 0.01\, m_a$, and the coupling parameter is $F=0.1$. The dashed lines show the analytical estimates, while the points correspond to individual numerical solutions to the field equations.}
\label{fig:xivY2}
\end{figure}

\begin{figure}
\centering
\includegraphics[width=\linewidth]{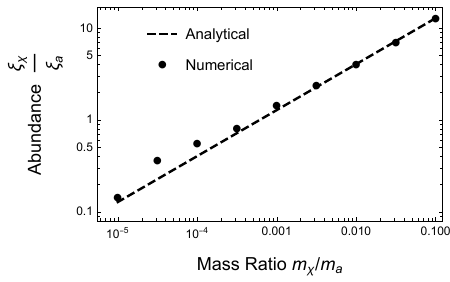}
\caption{The ratio of $\chi$ and axion abundances in the DPQ model, $\xi_{\chi}/\xi_{a}$, is plotted against the mass of the $\chi$ field, showing that $\xi_\chi\propto\sqrt{m_\chi}$. The axion mass is fixed at $m_a=10^{-10} \mbox{ eV}$, the initial field value is $Y_i=\chi_i/S=5$, and the coupling parameter is $F=0.1$. The dashed line shows the analytical estimates, while the points correspond to individual numerical solutions to the field equations.}
\label{fig:xivm1}
\end{figure}

\begin{figure}[t]
\centering
\includegraphics[width=\linewidth]{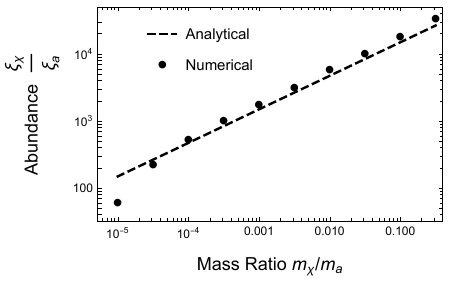}
\caption{The ratio of $\chi$ and axion abundances in the IPQ model, $\xi_{\chi}/\xi_{a}$, is plotted against mass of the $\chi$ field, showing that $\xi_\chi\propto\sqrt{m_\chi}$. The axion mass is fixed at $m_a=10^{-10} \mbox{ eV}$, the initial field value is $Y_i=\chi_i/S=5$, and the coupling parameter is $F=0.1$. The dashed line shows the analytical estimates, while the points correspond to individual numerical solutions to the field equations.}
\label{fig:xivm2}
\end{figure}

\begin{figure}
\centering
\includegraphics[width=\linewidth]{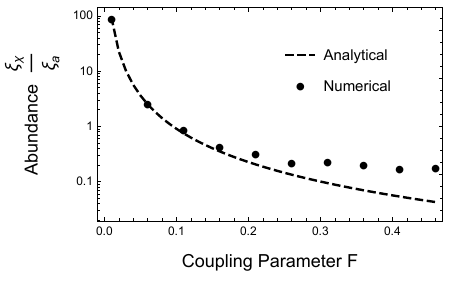}
\caption{The ratio of $\chi$ and axion abundances in the DPQ model, $\xi_{\chi}/\xi_{a}$, is plotted against the coupling parameter $F$, showing roughly that $\xi_\chi\propto F^{-2}$. The axion mass is fixed at $m_a=10^{-10} \mbox{ eV}$, the $\chi$ mass is $m_\chi = 5\times 10^{-4}\, m_a$, and the initial field value is $Y_i=\chi_i/S=5$. The dashed line shows the analytical estimates, while the points correspond to individual numerical solutions to the field equations.}
\label{fig:xivF1}
\end{figure}

\begin{figure}[t]
\centering
\includegraphics[width=\linewidth]{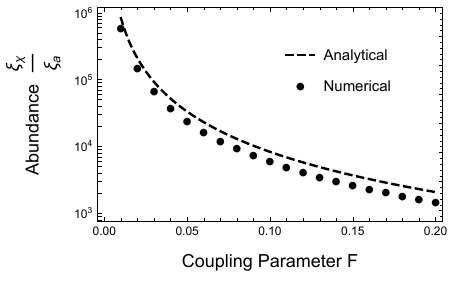}
\caption{The ratio of $\chi$ and axion abundances in the IPQ model, $\xi_{\chi}/\xi_{a}$, is plotted against the coupling parameter $F$, showing roughly that $\xi_\chi\propto F^{-2}$. The axion mass is fixed at $m_a=10^{-10} \mbox{ eV}$, the $\chi$ mass is $m_\chi = 0.01\, m_a$, and the initial field value is $Y_i=\chi_i/S=5$. The dashed line shows the analytical estimates, while the points correspond to individual numerical solutions to the field equations.}
\label{fig:xivF2}
\end{figure}

The initial field value $Y_i$ controls the abundance of the axion and the abundance of the $\chi$ field. 
Figures~\ref{fig:xivY1} and~\ref{fig:xivY2} show the ratios of abundances to the standard axion abundance $\xi/\xi_{a,std}$ in the DPQ and IPQ models respectively. The figures show numerical computations of these abundances from evolving the field equations, as well as analytical estimates from Eq.s~\ref{xiax} and~\ref{xichi}, for varying choices of $Y_i$. In the lower panel of both Fig.~\ref{fig:xivY1} and fig.~\ref{fig:xivY2}, the ratio $\xi_\chi/\xi_a$ is shown, with the analytical estimate from Eq.~\ref{xichixia}.

The mass of the $\chi$ field determines its abundance and has little effect on the abundance of the axion. With the choices $Y_i=5$ and $F=0.1$, Fig.s~\ref{fig:xivm1} and~\ref{fig:xivm2} show the ratio $\xi_\chi/\xi_a$ computed numerically and analytically while varying the mass $m_\chi$, for the DPQ and IPQ models respectively.

Finally, the coupling parameter $F$ controls the strength of the interaction between the axion and the $\chi$ field, which effects the relic abundance of the $\chi$ field primarily. Stronger interaction strengths suppress $\xi_\chi$ more effectively, and vice versa. 
Fixing $Y_i$ and $m_\chi$, we show the effect of varying $F$ in the DPQ model in Fig.~\ref{fig:xivF1} and in the IPQ model in Fig.~\ref{fig:xivF2}.

\section{Particle Physics Constraints on New Scalar}\label{Constraints}

Since we have introduced a new light scalar $\chi$, we should check if it leads to new effects that should have already shown up in experiments. At late times, the axion and the $\chi$ field have relaxed to small values, at which point the following is the leading order interaction between the new field and the axion
\beq
\Delta\mathcal{L} = \pm {1\over2} F^2(\partial\theta)^2 \chi^2
\eeq
where we have expanded $f(\chi)$ around zero for small $\chi$.
Since the axion's couplings to the Standard Model are already highly suppressed, if this is the leading interaction, then the $\chi$ field will only interact with the Standard Model by processes that are even more suppressed, as it needs the axion as a mediator. Hence, there does not appear to be any direct constraints on $\chi$. 

On the other hand, one could introduce direct couplings between $\chi$ and the Standard Model, such as through $\chi^2 H^\dagger H$, where $H$ is the Higgs. These couplings are not required by any fundamental considerations, if $\chi$ emerges from some other sector, but may be interesting to consider in future work. 

\section{Conclusions}\label{Conclusions}

In this work, we have introduced a new mechanism that can alter the relic axion abundance. This is accomplished by coupling the heavy Peccei-Quinn field $\Phi$ to a new light scalar $\chi$ and obtaining an effectively dynamical PQ scale in the early universe. This new behavior is interesting because it modifies the cosmological dynamics of the axion. Although the early-time behavior is altered, in the late universe, the axions still behave as nonrelativistic matter, and still provide an excellent candidate for dark matter. The abundance can improve (or worsen) the fit to data, and so this dynamical PQ-scale mechanism can be useful in understanding the range of possibilities for various QCD axion realizations. As future work, it would be interesting to investigate to what extent this new field $\chi$ may fit in with explicit microscopic constructions, such as \cite{Kim1979,Shifman1980,Zhitnis1980,Dine1981}.

We have proposed two simple models for the dynamical PQ-scale mechanism. In the DPQ model, the PQ scale begins effectively larger than its late-time value, corresponding to an axion which has a smaller effective mass during the onset of oscillations compared to its mass at late times. This type of evolution results in an enhanced relic abundance for the axion, compared to a standard QCD axion with the same late-time mass, and can be interesting for models of the axion with motivation for a mass too heavy to produce enough dark matter via the standard misalignment mechanism. A specific example of precisely this type of heavy axion is found in the so-called ``horizontal branch hint" \cite{Giannotti2016}. This is an inference about cooling mechanisms in horizontal branch stars which may be explained with an axion-photon coupling. Importantly this requires axions to be too heavy to be the dark matter in standard models; so our altered model can improve this situation.

The second of the two models is the IPQ model, where the PQ scale begins effectively small and grows, and thus the axion mass is lighter at later times. This nontrivial evolution can relax the lower bound on the mass of the QCD axion, which is typically $m_a \gtrsim 10^{-5}$\,eV, as set by demanding that the axion not be produced in amounts large enough to overclose the universe (i.e. larger than the observed relic abundance of dark matter today).
Of particular interest are situations where the PQ scale is set by a well-motivated high energy scale, such as the GUT scale, which results in an axion mass much smaller than the standard lower bound ($m_a\sim 10^{-10}\mbox{ eV}$). To avoid overclosing the universe, the increasing-PQ-scale model can result in a decreased relic abundance for this axion compared to its standard misalignment result. Although a huge reduction in abundance for both fields is not easy to achieve, it suggests an avenue worth pursuing.
This may have profound implications for unification, and deserves to be examined further.

Throughout, we have also commented on some experimental searches for axion dark matter, and how their successes may be accommodated and explained in the context of our dynamical PQ scale mechanism.
For example, the IPQ model can bring an axion mass of $m_a\sim 10^{-6}$\,eV into agreement with the observed abundance, which is very interesting for the ADMX search.

We have provided numerical results that demonstrate the effects these two models can have, and further numerical work is needed to verify in detail whether these models can viably account for the dark matter in the parameter regimes suggested by current observations. Part of the further work should be to include inhomogeneities in the fields.
At this state, the simple models presented in this work illustrate that the dynamical PQ scale mechanism can indeed influence observable parameters of the axion, and therefore this mechanism can be useful in exploring a broader parameter space for the QCD axion. One can readily imagine additional models that modify the effective PQ scale via couplings other than those discussed in this paper. It remains an interesting question as to what the microscopic constructions of any these possibilities could be. An important issue to address is the plausibility of the lightness of the new $\chi$ field. Variations on this model, in which the mass of $\chi$ is protected by extending the $U(1)$ PQ symmetry, are possible and are currently being explored.
Indeed all these issues mentioned above deserve further examination.

\section*{Acknowledgments}
M.~P.~H and Y.~L acknowledge support from the Tufts Visiting and Early Research Scholars’ Experiences Program (VERSE).
M.~P.~H is supported in part by National Science Foundation grant PHY-2013953.

\bibliography{DynPQ1}

\end{document}